\documentclass [12pt]{article}
\usepackage {amssymb}
\usepackage {amsmath}

\begin{document}
\noindent Invited talk delivered at the Fifth International
Conference on Computing Anticipatory Systems: (CASYS'01), Li\`ege,
Belgium, 13-18 August 2001. To appear in {\it International
Journal of Computing Anticipatory Systems} (2002).

\bigskip\medskip

\begin{center}
\Large{\textbf{Submicroscopic Deterministic Quantum Mechanics}}
\end{center}

\bigskip

\begin{center}
{\bf Volodymyr Krasnoholovets}
\end{center}

\bigskip

\begin{center}
\textit{Institute of Physics, National Academy of Sciences} \\
\textit{Prospect Nauky 46, UA-03028 Ky\"{\i}v, Ukraine} \\
\textit{Fax:} +(380 44) 265 1589; \  {\it E-mail:}
krasnoh@iop.kiev.ua
\\ {\it Web Page}: http://inerton.cjb.net
\end{center}

\bigskip

{\small{ {\bf Abstract.} So-called hidden variables introduced in
quantum mechanics by Louis de Broglie and David Bohm have been
revived in the recent works by the author. The start viewpoint was
the following: All the phenomena, which we observe in the quantum
world, should reflect structural properties of the real space.
Thus the scale $10^{-28}$ cm at which fundamental interactions
intersect has been treated as the size of a building block of the
space. The mechanics of a moving particle that has been
constructed is deterministic by its nature and shows that the
particle interacts with cells of the space creating elementary
excitations called "inertons" in a cellular substrate.  The
existence of inertons has been verified experimentally.

\medskip

 }}

\medskip
\bigskip

\section{Conceptual difficulties of quantum theory}

\hspace*{\parindent}

The main original physical parameters of quantum theory are
Planck's constant $h$ and de Broglie's wavelength $\lambda$. These
two enter into the two major quantum mechanical relationships for
a particle proposed by Louis de Broglie (see, e.g. de Broglie,
1986)
\begin{equation}
E=h\nu; \quad \quad \quad \lambda=h/p. \label{1}
\end{equation}
De Broglie believed that $E$ and $p$ were the energy and the
momentum of the particle, $\nu$ was the peculiar particle's
frequency that coincided with the frequency of a wave that
specified by the wavelength $\lambda$ and traveled together with
the particle. Later when Schr\"odinger's equation appeared and
Heisenberg proposed the uncertainty relations, the interpretation
of the said characteristics changed. Namely, the notion of the
particle was transformed to a "particle-wave" and hence ë and í
became characteristics of the particle-wave. Born interpreted the
square of the absolute magnitude of the wave $\psi$-function of
the Schr\"odinger equation as the probability of particle location
in a place described by the radius vector \textbf{r}. Thus Born
finally rejected any physical interpretation from a set of
parameters that described a quantum system. Since the end of the
1920s, only one parameter has been perceived as pure physical -
the particle-wave wavelength $\lambda$ called the de Broglie
wavelength. In experimental physics, those waves received also
another name -- the matter waves. Such a name directly says that
corpuscles (but not dim "particle-waves") are able to manifest a
wave behavior.

Since 1952 de Broglie followed two papers by Bohm (1952) (see also
Bohm, 1996) turned back to his initial ideas on the foundations of
the wave mechanics of particles. De Broglie (see, e.g. de Broglie,
1960, 1987) believed that a submicroscopic medium interfered with
the motion of a particle and the appropriated wave guided the
particle. He firmly believed the causal interpretation of quantum
mechanics and warned that the resolution of the issue should not
be based on the wave $\psi$-function formalism, as the
$\psi$-function was determined only in the phase space but not in
a real one. His own attempts were aimed at seeking for the form of
the so-called double solution.

In the case of the Dirac formalism things get worse. The formalism
introduced new additional notions such as spinors and Dirac's four-row
matrices, which allowed the calculation of the energy states of the quantum
system studied and changes in the states due to the influence of outside
factors. However, the formalism did not propose any idea on the reasons of
the wave behavior of matter and a nature of the particle spin.

So far, modern studies devoted to the foundations of quantum
mechanics have tried to reach the deepest understanding of quantum
theory reasoned that just the $\psi$-function formalism is
original and it is often exploited even on the scale of Planck
length $\sqrt{G \hbar /c^3} \sim 10^{-33}$ cm. This is especially
true for quantum field theory including quantum gravity (see, e.g.
Wallace, 2000; Sahni and Wang, 2000). Besides, there are views
that a gravitationally induced modification to the de Broglie's
wave-particle duality is needed when gravitational effects are
incorporated into the quantum measurement process (Ahluwalia,
1994, 2000; Kempf et al., 1995). Other approaches try to introduce
a phenomenological description based on the metric tensor g$_{ij}$
in typical quantum problems ('t Hooft, 1998). Classical Einstein
gravity is also exploited in condensed matter: some parameters
such as mass, spin, velocity, etc. are combined to provide an
effective "metric" that then is entered into the quantum
mechanical equations (e.g. Danilov et al. (1996) and Leonhard and
Piwnicki (1999)).

Thus, the trend has been forward to the entire intricacy: the
formalism of $\psi$-function penetrates to the Planck length
interior and the Einstein metric formalism advances to the same
scale as well. Nobody wishes to accept the fact that on the size
comparable with the de Broglie wavelength $\lambda$ of an object
methods of general relativity fail. No one wants to go deeply into
de Broglie's remark that the $\psi$-function is only a reflection
of some hidden variables of a particle moving in the real physical
space. The $\psi$-function is not the mother of particle nature
and therefore it cannot serve as a variable of the expansion of a
particle's characteristic in terms of $\psi$ at the size less than
the particle's de Broglie wavelength $\lambda$.

\section{New understanding}

\hspace*{\parindent}

Among new approaches describing gravity in the microworld, we can notice the
mathematical knot theory (see, e.g. Pullin, 1993), which has been developed
(Wallace, 2000) attempting to find rules to establish when one knot can be
transformed into another without untying it. In the theory, the question is
reduced to a certain knot invariant problem, which does not change with knot
deformations; knot invariants being deformed constantly by gauge
transformations should stay unchangeable. The approach is similar in many
aspects to concepts elaborated in elementary particle physics.

Of special note is the approach proposed by Bounias (1990, 2000)
and Bounias and Bonaly (1994, 1996, 1997). Basing on topology and
set theory, they have demonstrated that the necessity of the
existence of the empty set leads to the topological spaces
resulting in a "physical universe". Namely, they have investigated
links between physical existence, observability, and information.
The introduction of the empty hyperset has allowed a preliminary
construction of a formal structure that correlates with the
degenerate cell of space supporting conditions for the existence
of a universe. Besides, among other results we can point to their
very promising hypothesis on a non-metric topological distance as
the symmetric difference between sets: this could be a good
alternative to the conventional metric distance which so far is
still treated as the major characteristic in all concepts employed
in gravitational physics, cosmology, and partly in quantum
physics.

In my own line of research I started from the fact that on the
scale $\sim 10^{-28}$ cm constants of electromagnetic, weak and
strong interactions as functions of distance between interacting
particles intersect (see, e.g., Okun, 1988). On the other hand, in
the high energy physics theorists deal with an abstract
"superparticle" whose different states are electron, muon, quark,
etc. (see, e.g. Amaldi, 2000). A simple logical deduction suggests
by itself: the physical space at the said range has a peculiarity
that could be associated with presence of structural blocks which
one can call just superparticles (or elementary cells, or balls).
Then one may expect that a theory of the physical space densely
packed with those superparticles will be able to overcome many
difficulties which are insuperable in formal theories of both
quantum gravity and high energy physics. Thus a submicroscopic
theory being based on the structure of fine-grained space will be
able to widely expand our knowledge about the origin of matter,
the foundations of quantum mechanics and the foundations of
quantum gravity.

The first step of the theory, Krasnoholovets and Ivanovsky (1993),
Krasnoholovets (1997, 2000a, 2000b), focused on the appearance of
a particle from a superparticle, which initially was found in the
degenerate state. The particle has been defined as a local
curvature, or a local deformation of a superparticle and hence the
appearance of the deformation in a superparticle means the
induction of mass in it, $m \propto C{\kern 0.6pt}V_{\rm
sup}/V_{\rm part}$ ($C$ is the dimensional constant, $V_{\rm sup}$
is the initial volume of a degenerate superparticle and $V_{\rm
part}$ is the volume of the deformed superparticle, i.e. the
volume of the created particle). So the real space was regarded
rather as a substrate, or "quantum aether", and the notion of a
particle in it was adequately determined.

In condensed matter, we meet the effect of the deformation of the
crystal lattice in the surrounding of a foreign particle and the
solvation effect in liquids. Therefore, the second step of the
theory was the proposition that around a particle a deformation
coat was induced. This coat should play the role of a screen
shielding the particle from the degenerate space substrate. Within
the coat, the space substrate should be considered as a crystal
and superparticles here feature mass. Thus the coat may be treated
as a peculiar crystallite. The size of the crystallite was
associated with the Compton wavelength of the particle,
$\lambda_{{\kern 1pt}\rm Com} = h/mc$.

The next step needed a correct physical model of the motion of the
particle. From the solid state physics we know that the motion of
particles is accompanied with the motion of elementary excitations
of some sort, namely, the particle when it moves in a solid emits
and absorbs quasi-particles such as excitons, phonons, etc. By
analogy, the motion of the physical "point" (particle cell) in the
entirely packed space must be accompanied by the interaction with
surrounding coming "points" of the space, i.e. superparticle
cells. Hence the particle is scattered by structural blocks of the
space that in turn should lead to the induction of elementary
excitations in superparticles, which contact the moving particle.
The corresponding excitations were called "inertons" as the notion
"inertia" means the resistance to the motion (thus particle's
inertons reflect resistance on the side of the space in respect to
the moving particle). Each inerton carries a bit of the particle
deformation, that is, an inerton is characterized by the mass as
well. An inerton migrates from superparticle to superparticle by
relay mechanism. The deformation coat, or crystallite (by analogy
with crystal physics), is pulled by the particle: superparticles,
which form the crystallite, do not move from their positions in
the space substrate, however, the massive state of crystallite's
superparticles is passed on from superparticles to superparticles
along the whole particle path.

\bigskip

\section{Submicroscopic mechanics}

\hspace*{\parindent}

The Lagrangian that is able to satisfy the described motion of a particle
and the ensemble of its inertons can be written as (Krasnoholovets and
Ivanovsky, 1993)
\begin{equation}\label{2}
\begin{array}{l}
L= \frac 12 g_{ij} \frac {dX^{i}}{dt} \frac {dX^{j}}{dt} + \frac
12 \sum\limits_s g^{(s)}_{ij} \frac {dx_{(s)}^i}{dt_{(s)}} \frac {
dx_{(s)}^j}{dt_{(s)}}
\\
\quad\quad\quad  - \sum\limits_s \delta_{t-\Delta t_{(s)}, t_{(s)}
}{\kern 2pt} \frac {\pi}{T_{(s)}} {\kern 1pt}\Big[ X^i \sqrt {
g_{iq}{\kern 1pt}\theta {\kern 2pt} {\tilde g}_{(s)qj}}{\kern 4pt}
\frac {dx_{(s)}^j}{dt_{(s)}} + ({\bf v}_0)^{i} \sqrt { g_{iq}
{\kern 1pt} \theta {\kern 1pt} {\tilde g}_{(s)qj} }{\kern 4pt}
x_{(s)}^j \Big]
\end{array}
\end{equation}

\noindent where the first term characterizes the kinetics energy
of the particle, the second term characterizes the kinetics energy
of the ensemble of $N$ inertons, emitted from the particle and the
third term specifies the contact interaction between the particle
and its inertons. $X^i$ is the $i$th component of the position of
the particle; $g_{ij}$ is metric tensor components generated by
the particle; $({\bf v}_0)^i$ is the $i$th component of the
initial particle's velocity vector ${\bf v}_0$. Index s
corresponds to the number of respective inertons; $x_{(s)}^j$ is
the component of the position of the $s$th inerton; $g_{(s)ij}$ is
the metric tensor components of the position of the $s$th inerton.
$1/T_{(s)}$ is the frequency of collisions of the particle with
the $s$th inerton. Kronecker's symbol $\delta_{t-\Delta t_{(s)},
t_{(s)}}$ provides the agreement of proper times of the particle
$t$ and the $s$th inerton $t_{(s)}$ at the instant of their
collision ($\Delta t_{(s)}$ is the time interval after expire of
which, measuring from the initial moment $t = 0$, the moving
particle emits the $s$th inerton). The interaction operator $\sqrt
{ g_{iq} {\kern 2pt}\theta {\kern 2pt}{\tilde g}_{(s)qj}}$
possesses special properties: $\theta = 0$ during a short time
interval $\delta t$ when the particle and the $s$th inerton is in
direct contact and $\theta = 1$ when the particle and the $s$th
inerton fly apart along their own paths. Note that in the model
presented the metric tensor characterizes changing in sizes of the
particle and superparticles.

In the so-called relativistic case when the initial velocity ${\bf
v}_0$ of the particle is close to the speed of light $c$, the
relativistic mechanics prescribes the Lagrangian
\begin{equation}
L_{\rm rel} = -M_0c^2  \sqrt {1-{\bf v}_0^2/c^2}. \label{3}
\end{equation}
On examination of the relativistic particle, we shall introduce
into the Lagrangian (3) terms, which describe inertons and their
interaction with the particle. For this purpose, the following
transformation in (3) should be made (Krasnoholovets, 1997)

\begin{equation}\label{4}
\begin{array}{l}
L_{\rm rel} = - gc^2 \Big\{ 1- \frac 1{gc^2} \Big [ g_{ij} \frac {
dX^i}{dt} \frac {dX^j}{dt} + \sum\limits_s g^{(s)}_{ij} \frac {
dx_{(s)}^i}{dt_{(s)}} \frac {dx_{(s)}^j}{dt_{(s)}}         \\
\quad\quad\quad\quad\quad\quad - \sum\limits_{s} \delta_{t-\Delta
t_{(s)}, t_{(s)} } {\kern 2pt} \frac {\pi}{T_{(s)}} {\kern 1pt}
\Big( X^i \sqrt{ g_{iq}{\kern 2pt} \theta {\kern 2pt}{\tilde
g}_{(s)qj}} {\kern 3pt}\frac { dx_{(s)}^j}{dt_{(s)}} + ({\bf
v}_0)^i \sqrt{ g_{iq} {\kern 2pt} \theta {\kern 2pt}{\tilde
g}_{(s)qj}} {\kern 3pt} x_{(s)}^j \Big) \Big] \Big\}
\end{array}
\end{equation}

\noindent where $g = g_{ij}{\kern 1pt} \delta^{ij}$.

The Euler-Lagrange equations
\begin{equation}
\frac {d}{d{\kern 1pt}t_{(s)}} {\kern 2pt} \frac {\partial {\cal
L}}{\partial (d{\kern 0.5pt}Q/d{\kern 0.5pt}t_{(s)})} - \frac
{\partial{\cal L}}{\partial Q } = 0 \label{5}
\end{equation}

\noindent
written for the particle ($Q = X^i$) and the $s$th
inerton ($Q = x_{(s)}^i$) coincide for the Lagrangians ${\cal L}
=L$ (3) and ${\cal L} = L_{\rm rel}$ (4). This is true only
(Dubrovin et al. (1986)) in the case when the time $t$ entered
into the Lagrangians (3) and (4) is considered as the natural
parameter, i.e. $t = l/v_0$ where $l$ is the length of the
particle path.

For the variables $X_{(s)}^k= X^k
 (t_{(s)}$) and $x_{(s)}^k =
x^k(t_{(s)}$) one obtains from eq. (5) the equations of extremals
(written as functions of the proper time $t_{(s)}$ of the emitted
$s$th inerton):

\begin{equation}\label{6}
\frac {d^{{\kern 1pt}2} X_{(s)}^k}{d{\kern 1pt}t_{(s)}^2} + \Gamma
^k_{ij} \frac {dX_{(s)}^i} {d{\kern 1pt}t_{(s)}} \frac
{dX_{(s)}^j} {d{\kern 1pt}t_{(s)}} + \frac {\pi} {T_{(s)}} {\kern
2pt} g^{ki} {\kern 1pt}\sqrt { g_{iq} {\kern 2pt} \theta {\kern
2pt} {\tilde g}_{(s)qj} } {\kern 4pt}\frac {d {\kern
0.6pt}x_{(s)}^j}{d{\kern 1pt}t_{(s)}} = 0;
\end{equation}
\begin{equation}\label{7}
\frac {d^{{\kern 1pt}2} x_{(s)}^k}{d{\kern 1pt}t_{(s)}^2} +
{\tilde \Gamma}^k_{(s)ij} \frac{dx_{(s)}^i}{d{\kern 1pt}t_{(s)}}
\frac {d x_{(s)}^j}{d{\kern 1pt} t_{(s)}} - \frac {\pi} {T_{(s)}}
\Big [ g_{(s)}^{ki} {\kern 1pt}\sqrt {g_{iq}{\kern 2pt} \theta
{\kern 2pt}_{(s)qj} } {\kern 3pt}\Big( \frac {dX_{(s)}^j}{d{\kern
1pt}t_{(s)}} - ({\bf v}_0)^j \Big) \Big ] = 0;
\end{equation}

\noindent here, $\Gamma^k_{ij}$ and ${\tilde \Gamma}^k_{(s)ij}$
are symmetrical connections (see, e.g. Dubrovin et al., 1996) for
the particle and for the $s$th inerton, respectively; indices $i,
\, j, \, k$ and $q$ take values 1, 2, 3. When the particle and the
$s$th inerton adhere, the operator $\theta = 0$ and therefore the
termwise difference between eqs. (6) and (7) becomes

\begin{equation}\label{8}
\Big( \frac {d^{{\kern 1pt}2}X_{(s)}^k}{d{\kern 1pt}t_{(s)}^2} -
\frac {d^{{\kern 1pt}2} x_{(s)}^k}{d{\kern 1pt}t_{ (s)}^2} \Big) +
\Big( \Gamma^k_{ij} \frac {dX_{(s)}^i}{d{\kern 1pt}t_{(s)}} \frac
{ dX_{(s)}^j}{d{\kern 1pt}t_{(s)}} - {\tilde
\Gamma}^k_{(s)ij}\frac {dx_{(s)}^i}{d{\kern 1pt}t_{(s)}} \frac
{dx_{(s)}^j}{d{\kern 1pt}t_{(s)}} \Big) = 0.
\end{equation}
Eq. (8) specifies the merging the particle and the $s$th inerton
into a common system. This means the acceleration that the
particle experiences, coincides with that of the $s$th inerton.
Then the difference in the first set of parentheses in eq. (8) is
equal to zero and instead of eq. (8) we get

\begin{equation}\label{9}
\Gamma^k_{ij} {\kern 1pt}\frac {d X_{(s)}^i}{d{\kern 1pt}t_{(s)}}
\frac {d X_{(s)}^j}{d{\kern 1pt}t_{(s)}} = {\tilde
\Gamma}^k_{(s)ij}{\kern 1pt} \frac {dx_{(s)}^i}{d{\kern
1pt}t_{(s)}} \frac {d x_{(s)}^j}{d{\kern 1pt}t_{(s)}}.
\end{equation}

Coefficients $\Gamma ^k_{ij}$ and ${\tilde \Gamma}^k_{(s)ij}$ are
generated by the particle mass $M$ and the $s$th inerton mass
$m_{(s)}$, respectively, and that is why $\Gamma^k_{ij}/{\tilde
\Gamma}^k_{(s)ij} = M/m_{(s)}$. This signifies that relationship
(9) can be rewritten explicitly

\begin{equation}\label{10}
M v_{0s}^2 = m_{(s)}c^2
\end{equation}

\noindent for diagonal metric components of the particle and
inerton velocities, ($v_{0(s)}$ is the velocity of the particle
after its scattering by the $s$th inerton with initial velocity
$c$).

When the particle and the $s$th inerton bounce apart, we must
solve the total equations of motion, (6) and (7), i.e. all terms
in the equations should be held. However, if we allow the metric
tensors to be constant, the equations of motion may be simplified
to the form that does not include the second nonlinear term in
both eqs. (6) and (7). The structure and properties of the metric
tensors can be chosen as follows

\begin{equation}\label{11}
\begin{array}{l}
g_{ij}= \delta_{ij}M; \quad\quad\quad\quad    g^{ij}=
\delta^{ij}{\kern 1pt}/M; \quad\quad\quad   g_{ki}{\kern 2pt}
g^{iq} = \delta_{k}^{q}; \quad\quad                  \\ {\tilde
g}_{(s)ij}= \delta_{ij}{\kern 1pt} m_{(s)}; \quad \quad {\tilde
g}_{(s)}^{ij}= \delta ^{ij}/m_{(s)}; \quad\quad {\tilde g}_{(s)ki}
{\kern 2pt}{\tilde g}^{(s)iq} = \delta^{q}_{k}.
\end{array}
\end{equation}

Thus having given $g_{ij}$ and $g_{(s)ij}$ are equal to constant,
the second term in both eqs. (7) and (8) is made to be reduced to
zero. Relationships (11) and (10) allow transforming of the
interaction operator in eqs. (7) and (8) to forms

\begin{equation}\label{12}
g^{ki} {\kern 1pt} \sqrt{g_{iq}{\kern 2pt}\theta{\kern 2pt}{\tilde
g}_{(s)qj}} {\kern 1pt} \longrightarrow {\kern 1pt}\sqrt{\frac
{m_{(s)}}M} = \frac {v_{0(s)}^{k}}c;
\end{equation}
\begin{equation}\label{13}
 g^{ki} {\kern 1pt} \sqrt {g_{iq}{\kern 2pt} \theta {\kern 2pt}{\tilde
 g}_{(s)qj}}
{\kern 1pt}\longrightarrow {\kern 1pt}\sqrt{\frac M{m_{(s)}}} =
\frac c {v_{0(s)}^{k}}
\end{equation}
where $v_{0(s)}^{k}$ is the $k$th component of the vector ${\bf
v}_{0(s)}$. Thus expressions (12) and (13) permit the
transformation of eqs. (7) and (8) (in which second terms are
dropped) to the form

\begin{equation}\label{14}
\frac {d^{2} X_{(s)}^{k}}{d{\kern 1pt}t_{(s)}^{2}} + \frac {\pi
v_{0(s)}^{k}}{c{\kern 0.5pt}T_{(s)}} {\kern 2pt}\frac
{dx_{(s)}^{j}}{d{\kern 1pt}t_{(s)}} = 0;
\end{equation}
\begin{equation}\label{15}
\frac{d^{2} x_{(s)}^{k}}{d{\kern 1pt}t_{(s)}^{2}} - \frac{\pi c}
{v_{0(s)}^{k} T_{(s)}}{\kern 2pt} \Big[ \frac
{dX_{(s)}^{k}}{d{\kern 1pt}t_{(s)}} - (v_{0(s)})^{k} \Big] = 0.
\end{equation}

Initial conditions are
   $$
\frac {dX_{(s)}(t_{(s)} + \Delta t_{(s)}) \big|_{t(s)=0}} {d{\kern
1pt}t_{(s)}} = \frac {dX_{(s)}(\Delta t_{(s)})}{d{\kern
1pt}t_{(s)}} = v_{(s)0};
   $$
   $$
x_{(s)}\big|_{t(s)=0} = 0; \quad\quad\quad \frac{dx_{(s)}}{d{\kern
1pt}t_{(s)}}\big|_{t(s)=0} = c.
   $$

If we consider the ensemble of inertons as the whole object, an
inerton cloud with the rest mass $m_0$, which surrounds a moving
particle with the rest mass $M_0$ then the Lagrangian may be
presented as

\begin{equation}\label{16}
L= - M_0{\kern 1pt} c^2 \Big\{ 1- \big( \frac 1 {M_0{\kern
1pt}c^2} \big) \Big[ M_0 \big( \frac{d X} {d{\kern 1pt} t} \big)^2
+m_0 \big( \frac {d x}{d {\kern 1pt} t} \big)^2 - \frac {2 \pi } T
\sqrt{M_0 m_0}{\kern 1pt} \Big( X \frac {d x} {d {\kern 1pt} t} +
v_0 {\kern 1pt} x \Big) \Big] \Big\}^{1/2}.
\end{equation}

Thus the particle moves along the $X$-axis with the velocity
$d{\kern 0.5pt}X/d{\kern 1pt}t$ ($v_{0}$ is the initial velocity);
$x$ is the distance between the inerton cloud and the particle,
$d{\kern 0.5pt}x/d{\kern 1pt}t$ is the velocity of the inertons
cloud, and $1/T$ is the frequency of collisions between the
particle and cloud. The equations of motion are reduced to the
following

\begin{equation}\label{17}
\frac {d^{\kern 1pt 2} X }{d{\kern 1pt}t^2} + \frac {\pi}{T}{\kern
1pt}\frac {v_0}{c} {\kern 1pt} \frac {d{\kern 1pt} x} {d{\kern
1pt} t} = 0;
\end{equation}
\begin{equation}\label{18}
\frac {d^2 x}{d{\kern 1pt}t^2}- \frac {\pi}{T}{\kern 1pt
}\frac{c}{v_0}{\kern 1pt} \Big( \frac {d X}{d {\kern 1pt}t} - v_0
\Big) = 0.
\end{equation}

The corresponding solutions to eqs. (17) and (18) for the particle
and the inerton cloud are

   $$
\frac {d X}{d{\kern 1pt}t} = v_0 \cdot \big(1- \big|\sin \frac
{\pi t}{T} \big| \big);
   $$
\begin{equation}\label{19}
X(t) = v_0{\kern 1pt} t + \frac \lambda \pi \cdot \Big\{ (-1)^{[
\frac t T ]}{\kern 1pt} \cos \frac {\pi t}{T}  - \big( 1 + 2{\kern
1pt} \big[ \frac t T \big] \big) \Big\};
\end{equation}
    $$
    \lambda= v_0 {\kern 1pt} T;
    $$
    $$
x = \frac \Lambda \pi {\kern 2pt}\big|\sin \frac {\pi t}{T} \big|;
\\
    $$
\begin{equation}\label{20}
\frac {d x}{d{\kern 1pt}t} = c{\kern 1pt}(- 1)^{[\frac tT]} {\kern
1pt}\cos \frac {\pi t} {T};
\end{equation}
    $$
\Lambda = c {\kern 2pt} T.
    $$
Expressions (19) show that the velocity of the particle
periodically oscillates and $\lambda$ is the amplitude of
particle's oscillations along its path. In particular, $\lambda$
is the period of oscillation of the particle velocity that
periodically changes between $v_0$ and zero. The inertons cloud
periodically leaves the particle and then comes back; $\Lambda$ is
the amplitude of oscillations of the cloud.

The frequency of collisions of the particle with the inerton cloud
allows the presentation in two ways: 1) via the collision of the
particle with the cloud, i.e., $1/T = v_{0}/\lambda$ and 2) via
the collision of the inerton cloud with the particle, i.e., $1/T =
c/\Lambda$. These two expressions result into the relationship

\begin{equation}\label{21}
 \frac {v_0}{\lambda} = \frac c{\Lambda},
\end{equation}
which connects the spatial period $\lambda$ of oscillations of the
particle with the amplitude $\Lambda$ of the inertons cloud, i.e.,
maximal distance to which inertons are removed from the particle.

If we introduce a new variable

\begin{equation}\label{22}
\frac {d\varkappa}{d{\kern 1pt}t} = \frac {d x}{d{\kern 1pt}t} -
\frac {\pi} {T} {\kern 1pt} X {\kern 1pt} \sqrt{\frac {M_0}{m_0}}
\end{equation}
in the Lagrangian (16), we arrive to the canonical form on
variables for the particle

\begin{equation}\label{23}
L = - M_0{\kern 0.5pt}c^2 \Big\{ 1- \frac 1{M_0{\kern 0.5pt}c^2}
\Big[ M_0 \big( \frac {d X}{d{\kern 1pt}t}\big)^{2} - M_0 \big(
\frac {2\pi}{2T} \big)^2 X^2 + {\kern 2pt} m_0 \big(\frac {d
\varkappa}{d{\kern 1pt}t} \big)^2 - \frac {2 \pi}T {\kern 1pt} v_0
{\kern 1pt} x {\kern 1pt}\sqrt{\frac {M_0}{m_0} }{\kern 2pt} \Big]
\Big\}^{1/2}.
\end{equation}

This Lagrangian allow us to obtain (Krasnoholovets, 1997) the effective
Hamiltonian of the particle that describes its behavior relative to the
center of inertia of the particle-inerton cloud system

\begin{equation}\label{24}
H_{\rm eff} = \frac 12 {\kern 1pt}\frac {p^2}M + \frac 12 {\kern
1pt} M \big( \frac {2\pi}{2T} \big)^2 X^2
\end{equation}
where $M = \frac {M_0}{\sqrt {1-v_0^2/c^2}}$ (and also $m = \frac
{m_0}{\sqrt{1-v_0^2/c^2}}$). The harmonic oscillator Hamiltonian
(24) allows one to write the Hamilton-Jacobi equation for a
shortened action $S_{1}$ of the particle
\begin{equation}\label{25}
\frac 1{2M} \Big( \frac {\partial S_1}{X} \Big)^2 + \frac 12 M
\big( \frac {2\pi}{2T} \big )^2 X^2 = E.
\end{equation}
Here $E$ is the energy of the moving particle. Introduction of the
action-angle variables leads to the following increment of the
particle action within the cyclic period $2T$ (Krasnoholovets and
Ivanovsky, 1993)
\begin{equation}\label{26}
\Delta S_1 = \oint  p{\kern 1pt}{\kern 2pt} d{\kern 0.5pt} X = E
\cdot 2{\kern 0.5pt}T.
\end{equation}
 One can write Eq. (26) via the frequency $\nu = 1/2T$ as well. At
the same time $1/T$ is the frequency of collisions of the particle
with its inertons cloud. Owing to the relation $E = \frac 12
Mv_0^2$ we also get
\begin{equation}\label{27}
\Delta S_1 = Mv_0 \cdot v_0 T = p_{{\kern 1pt}0} {\kern
1pt}\lambda
\end{equation}
where $p_{\kern 1pt 0} = Mv_0$ is the particle initial momentum.
Now if we equate the values $\Delta S_1$ and Planck's constant
$h$, we obtain instead of expressions (26) and (27) major
relationships (1), which form the basis of conventional quantum
mechanics.

De Broglie (1986), when writing relationships (1), noted that they
resulted from the comparison of the action of a particle moving
rectilinearly and uniformly (with the energy $E$ and the momentum
$Mv_0$) and the phase of a plane monochromatic wave extended in
the same direction (with the frequency $E/h$ and the wavelength
$h/{M v_0}$). Yet the first relation in (1) he considered as the
main original axiom of quantum theory.

In our case, expressions (26) and (27) have been derived starting
from the Hamiltonian (24) or the Hamilton-Jacobi eq. (25) of the
particle. The main peculiarity of our model is that the
Hamiltonian and the Hamilton-Jacobi equation describe a particle
whose motion is not uniform but oscillatory. It is the space
substrate, which induces the harmonic potential responding to the
disturbance of the space by the moving particle. The oscillatory
motion of the particle is characterized by the relation
\begin{equation}\label{28}
\lambda = v_0 T
\end{equation}
which connects the initial velocity of the particle $v_0$ with the
spatial period of particle oscillations $\lambda$ (or the free
path length of the particle) and the time interval $T$ during
which the particle remains free, i.e. does not collide with its
inerton cloud. On the other hand, relation (28) holds for a
monochromatic plane wave that spreads in the real physical space:
$\lambda$ is the wavelength, $T$ is the period and $v_0$ is the
phase velocity of the wave. Thus with the availability of the
harmonic potential, the behavior of the particle follows the
behavior of a wave and, therefore, such a motion should be marked
by a very specific value of the adiabatic invariant, or increment
of the particle action $\Delta S_1$ within the cyclic period. It
is quite reasonable to assume that in this case the value of
$\Delta S_1$ is minimum, which is equal to Planck's constant $h$.

Two relationships (1) immediately allow the deduction of the
Schr\"odinger equation (de Broglie, 1986). Moreover, the presence
of the proper time of a particle in the Schr\"odinger equation
(Krasnoholovets, 1997) signifies that the equation is Lorentz
invariant. The wave $\psi$-function acquires a sense of the
imaging of a real wave function that characterizes the motion of a
complicated formation -- the particle and its inerton cloud. The
real wave function (and its wave $\psi$-function imaging or map)
is defined in the range that is exemplified by the dimensions of
the particle's inerton cloud: $\lambda$ along the particle path
and $2\Lambda$ in transversal directions. In such a manner,
inertons acquire the sense of a substructure of the matter waves
and should be treated as carriers of inert properties of matter.
Heisenberg's uncertainty relations gain a deterministic
interpretation as a quantum system now is complemented by the
inerton cloud; therefore, an unknown value of the momentum of the
particle automatically is compensated by the corresponding
momentum of the particle's inertons.

\bigskip

\section{Spin and relativistic approximation}

\hspace*{\parindent}
 The notion of spin of a particle is
associated with an intrinsic particle motion. Several tens of
works have been devoted to the spin problem. Major of them is
reviewed in the recent author's paper (Krasnoholovets, 2000a).
Here we add some recent references: Chashihin (2000), Rangelov
(2001), Danilov et al. (1996), Plyuschay (1989, 1990, 2000). Main
ideas of the works quoted in Krasnoholovets (2000a), and in the
mentioned references are reduced to a moving particle that is
surrounded by a wave, or a small massless particle, or an ensemble
of small massless particles, which engage in a circular motion.

Having tried the introduction of the notion of spin in the concept
presented, let us look at the situations in which the particle
spin manifest itself explicitly. First, it appears as an
additional member $\pm \frac h2$ to the projection onto the z-axis
of the moment of momentum ${\bf r} \times M {\bf v}_0$ of a
particle. Second, it introduces the correction $\pm \frac {e B_z
h}{2M}$ to the energy of a charged particle in the magnetic field
with the projection of the induction onto the $z$-axis equals to
$B_z$. Third, it provides for the Pauli exclusion principle.

Of course, it seems quite reasonable to assume that the spin in
fact reflects some kind of proper rotation of the particle.
However, we should keep in mind that the operation 'rotor' is
typical for the electromagnetic field that the particle generates
in the environment when it starts to move. In other words, the
appearance of the electromagnetic field in the particle
surrounding one may associate just with its proper rotation of
some sort. In our concept, superparticles that form the space net
are not rigid; they fluctuate and allow local stable and unstable
deformations. Thus the particle may be considered as not rigid as
well. In this case along with an oscillating rectilinear motion,
the particle is able to undergo some kind of an inner pulsation,
like a drop. Besides the pulsation can be oriented either along
the particle velocity vector or diametrically opposite to it. Then
the Lagrangians (16) and (23) change to the matrix form
(Krasnoholovets, 2000a)
\begin{equation}\label{29}
{\cal L}  = || {\cal L}_{\alpha}||, \quad\quad  \alpha = \uparrow,
\, \downarrow.
\end{equation}
The function ${\cal L}_{\alpha}$ can be written as
\begin{equation}\label{30}
{\cal L}_{\alpha} = - g{\kern 1pt}c^2 \sqrt{1 - \frac 1{g{\kern
1pt}c^2}{\kern 1pt}\big[ U_{(\rm {spat})} + U_{({\rm intr})\,
\alpha} \big]}.
\end{equation}
Here U$_{({\rm spat})}$ is the same as in expression (2) and
$U_{({\rm intr})\, \alpha}$ is similar to U$_{({\rm spat})}$,
however, all spatial coordinates (and velocities) are replaced for
the intrinsic ones: \ ${\bf X} \rightarrow {\vec \Xi}_{\alpha}$
for the particle and ${\bf x} \rightarrow {\vec \xi}_{\alpha}$ for
the inerton cloud. So inertons carry bits of the particle
pulsation as well. The intrinsic motion is treated as a function
of the proper time of the particle $t$. Then the equations of
motion and the solutions to them are quite similar to those
obtained in the previous section. The intrinsic velocity $d{\kern
1.5pt} \Xi_{\alpha}/d{\kern 1pt}t$ ranges between $ \pm {\kern
1pt} v_{0}$ ("+" if $\alpha = \uparrow$ and "--" if $\alpha =
\downarrow$) and zero; $d {\kern 1pt}\xi_{\alpha}/d{\kern 1pt}t$
ranges between $\pm c$ and zero within the segment $2\Lambda$ of
the spatial path of the inerton cloud. Such a motion is
characterized by relationships similar to (26) and (27) and hence
is marked by Planck's constant $h$.

The intrinsic variables do not appear in the case of a free moving particle.
However, an external field being superimposed on the system is able to
engage into the variables. Then we can write the wave equation for the spin
variable of the particle
\begin{equation}\label{31}
\Big( \frac {{\hat {\bf \Pi}}_{\alpha}^{\kern 0.6pt 2}}{2M} -
e_{\alpha} {\kern 1pt} \varepsilon_{\alpha} \Big) \chi_{\alpha} =
0
\end{equation}
where the operator ${\hat {\bf \Pi}}_{\alpha} = ({\hat {\vec
\pi}}_{\alpha} - e {\bf A})$ and ${\hat {\vec \pi}}_{\alpha} =
-i\hbar {\kern 1.5pt}d/d{\kern 2pt} {\vec \Xi}_{\alpha}$ is the
operator of the intrinsic momentum of the particle, $e$ and ${\bf
A}$ are the electric charge and the vector potential of the field,
respectively. $\chi_{\alpha}$ is the eigenfunction and
$\varepsilon_{\alpha}$ is the eigenvalue; the function $e_{\alpha}
= 1$ if $\alpha =\uparrow$ and $e_{\alpha} = - 1$ if $\alpha
=\downarrow$.

If the induction of the magnetic field has only one component
$B_z$ aligned with the $z$-axis, the solution to eq. (31) becomes
\begin{equation}\label{32}
\varepsilon_{\alpha} = e_{\alpha}{\kern 1pt}\frac {e{\kern 0.6pt}
\hbar{\kern 0.6pt} B_{z}}{2M};
\end{equation}
\begin{equation}\label{33}
\chi_{\alpha} = \pi^{-1/4} \exp \big[-\frac {{\bf \pi}_{\alpha \,
x} - e A_x}{2{\kern 0.8pt} e{\kern 0.6pt}\hbar {\kern 0.6pt} B_z}
\big].
\end{equation}
So $\varepsilon_{\uparrow, \, \downarrow} = \pm {\kern 1.5pt}
\frac{e\hbar B_z}{2M}$ and therefore the eigenvalues of the
so-called spin operator ${\bf S}_{{\kern 1pt}\uparrow,
\downarrow}$ are
\begin{equation}\label{34}
S_{\uparrow, \downarrow \, \, z} =  \pm  \frac {\hbar} 2;
\quad\quad S_{\uparrow, \downarrow \, \, x} = S_{\uparrow,
\downarrow \, \, y} = 0.
\end{equation}

Thus, the intrinsic motion introduced above satisfies the behavior
of a particle in the magnetic field. The total orbital moment of
the electron in an atom includes the spin contribution proceeding
just from the interaction of the electron with an magnetic field.
Moreover, the availability of two possible antipodal intrinsic
motions of the particle allows the satisfaction of the Pauli
exclusion principle. Consequently, the model of the spin described
complies with the three said requirements.

Now the total Hamiltonian of a particle can be represented in the form of
\begin{equation}\label{35}
H_{\uparrow, \downarrow} = c {\kern 1pt} \sqrt{{\bf p}^2  + {\vec
\pi}^{{\kern 1.5pt}2}_{\uparrow, \downarrow} + M_0^2c^2}
\end{equation}
(a similar Hamiltonian describes the particle's inerton cloud). As
can easily be seen from expression (35), the spin introduces an
additional energy to the particle Hamiltonian transforming it to a
matrix form. Then following Dirac we can linearize the matrix
$H_{\uparrow, \downarrow}$ and doing so we will arrive to the
Dirac Hamiltonian
\begin{equation}\label{36}
{\hat H}_{\rm Dirac} = c {\kern 1.5pt}{\hat {\vec \alpha}}{\kern
1pt} {\hat{\vec p}} + \hat\rho_{\kern 0.5pt 3}M_0{\kern 1pt}c^2.
\end{equation}

At this point, information on the matric operators ${\hat {\vec
\pi}}_{\uparrow, \downarrow}$ goes into the Dirac matrices. Thus
from the physical point of view the Dirac transformation (36) is
substantiated only in the case when the initial Hamiltonian is a
matrix as well. And just this fact has been demonstrated in the
theory proposed.

The Dirac formalism is correct in the range $r \geq h/Mc$ and is
restricted by the amplitude of inerton cloud $\Lambda = \lambda
{\kern 1pt}c/v_0$. At $r < h/Mc$ the approach described above can
easily be applied. It has been pointed out (Krasnoholovets, 2000a)
that the inerton cloud and the oscillatory mode of the
crystallite's superparticles, which vibrate in the environment of
the particle, cause the nature of spinor components. Two possible
projections of spin enlarge the total number of the Dirac matrices
and the spinors to four.

The submicroscopic consideration allows one sheds light on the
interpretation of the so-called negative kinetic energy and the
negative mass of rest of a free particle, which enter into the
solutions of the corresponding Dirac equation (see, e.g. Schiff,
1959). The negative spectral eigenvalues $E_{-} = -\sqrt {c^2{\bf
p}^2 + m^2c^4}$ are interpreted as states with the negative energy
of the particle (and because of that Dirac proposed to refer it to
the energy of the positron). However, the presence of the inerton
cloud that oscillates near the particle lets us to construe the
eigenvalues of the particle as a spectrum of "left" and "right"
inerton waves which respectively emitted and absorbed by the
particle. Such waves, $\varphi_{\alpha \, \, -} =
\varphi_{\alpha}(r - at)$ and $\varphi_{\alpha \, +} =
\varphi_{\alpha}(r + at)$ where $\alpha$ specifies the spin
projection, depend on the space variables which can be made
identical, while the time variable $t$ is entered as either
$+{\kern 0.6pt}t$ or $-{\kern 0.3pt}t$. In quantum mechanics the
operator $i\hbar{\kern 2pt} {\partial}/{\partial {\kern 1pt} t}$
just corresponds to the particle energy $E$. Thus, we can
interpret the positive eigenvalue $E_{{\kern 0.5pt}+}$ as the
total energy of the inerton cloud that moves away from the
particle while the negative eigenvalue $E_{-}$ as the total energy
of the inerton cloud that comes back to the particle.

This must be paralleled with the recent research conducted by
Dubois (2000a,b), who has studied anticipation in physical systems
considering anticipation as their inner property which is embedded
in the system but is not a model-based prediction. In particular,
it has been shown by Dubois that such a property is inherent to
electromagnetism and quantum mechanics. Namely, Dubois (2000b)
started from space-time complex continuous derivatives which were
constructed in such a way that it gave the discrete forward and
backward derivatives $\partial^{{\kern 0.5pt}\pm} /
\partial {\kern 1pt} t$. Here, Dubois's methodology may be
justified in terms of the present submicroscopic approach because
the derivative $\partial^{+} / \partial {\kern 1pt} t$ might be
referred to inertons flying away from the particle and the
derivative $\partial^{-} / \partial {\kern 1pt} t$ might be
assigned to inertons moving backward to the particle. Besides the
two types of velocities are present in anticipatory physical
systems, so called "phase" and "group" velocities. These two
velocities would also be ascribed to two opposite flows of
inertons. We can also emphasize that the Dubois' idea about the
masses of particles as properties of space-time shifts is also
very close to the author's and Bounias (1990) hypothesis on mass
as a local deformation in the space net. Note that the hypothesis
has found future trends (Bounias and Krasnoholovets, 2002): it
allows evidence in terms of the topology and fractal geometry.

\bigskip

\section{Inertons in action: experimental verification}

\hspace*{\parindent}

$\underline{\S {\bf 1.}}$ The photoelectric effect occurring under
strong irradiation in the case that the energy of the incident
light is essentially smaller than the ionization potential of gas
atoms and the work function of the metal has been reconsidered
from the submicroscopic viewpoint. It has been shown
(Krasnoholovets, 2001a) that the (nonlinear) multiphoton theory,
which has widely been used so far, and the effective photon
concept should be changed for a new methodology. The author's
approach was based on the hypothesis that inerton clouds are
expanded around atoms' electrons. That means that the effective
cross-section $\sigma$ of an atom's electron together with the
electron's inerton cloud falls within the range between
$\lambda^2$ and $\Lambda^2$ (i.e. $10^{-16} \ {\rm cm}^{2} \ <
\sigma < 10^{-12}$ cm$^{2}$) that much exceeds the cross-section
area of the actual atom size, $10^{-16}$ cm$^{2}$. The intensity
of light in focused laser pulses used for the study of gas
ionization and photoemission from metals was of the order of
$10^{12}$ to $10^{15}$ W/cm$^{2}$. Thus several tens of photons
simultaneously should pierce the electron's inerton cloud and at
least several of them could be engaged with the cloud's inertons
and scattered by them. Consequently, the electron receives the
energy needed to release from an atom or metal. The theory indeed
has been successfully applied to the numerous experiments
(Krasnoholovets, 2001a).

\medskip
\bigskip

$\underline{\S {\bf 2.}}$ In condensed media, inerton clouds of
separate particles (electrons, atoms, and molecules) should
overlap forming the entire elastic inerton field, which densely
floods in the media. It has been theoretically shown
(Krasnoholovets and Byckov, 2000) that in this case the force
matrix W that determines branches of acoustic vibrations in solids
comprises of two members: $W = V_{\rm ac} + V_{\rm iner}$. Here
the first member is responsible for the usual elastic
electromagnetic interaction of atoms and is responsible just for
the availability of acoustic properties of solids, but the second
one is originating from the overlapping of atoms' inerton clouds.
It is remarkable that each of the members is involved in the
expression for $W$ equally. Therefore, an inerton wave striking an
object will influence the object much as an applied ultrasound.
Among the features of ultrasound, one can call destroying,
polishing, and crushing. It was anticipated that inerton waves
would act on specimens in a similar manner. A power source of
inerton waves is the Earth: any mechanical fluctuations in the
Earth should generate corresponding inerton waves. Two types of
inerton flows one can set off in the terrestrial globe. The first
flow is caused by the proper rotation of the Earth. Let A be a
point on the Earth surface from which an inerton wave is radiated.
If the inerton wave travels around the globe along the West-East
line, its front will pass a distance $L_{1} = 2\pi R_{{\kern 1pt}
\rm Earth}$ per circle. The second flow spreads along the
terrestrial diameter; such inerton waves radiated from A will come
back passing distance $L_{2} = 4R_{{\kern 0.5pt}\rm Earth}$. The
ratio is

\begin{equation}\label{37}
\frac {L_1}{L_2} = \frac {\pi}{2}.
\end{equation}
If in point A we locate a material object which linear sizes
(along the West-East line and perpendicular to the Earth surface)
satisfy relation (37), we will receive a resonator of the Earth
inerton waves.

We have studied specimens (razor blades) put into the resonator for several
weeks. By using the scanning electron microscope, in fact, we have
established difference in the fine morphological structure of cutting edge
of the razor blades while the morphologically more course structure remains
well preserved.

Note that the Earth inerton field is also the principal mover that
launched rather fantastic quantum chemical physical processes in
Egypt pyramids (Krasnoholovets, 2001b), power plants of the
ancients that has recently been proved by Dunn (1998).

\medskip
\bigskip

$\underline{\S {\bf 3.}}$ Just recently, the inerton concept has
been justified in the experiment on the searching for hydrogen
atoms clustering in the $\delta$-KIO$_3\cdot$HIO$_3$ crystal
(Krasnoholovets et al., 2001). It has been assumed that vibrating
atoms should induce the inerton field within the crystal. This in
turn should change the paired potential of interatomic
interaction. Taking into account such a possible alteration in the
potential, we have calculated the number of hydrogen atoms in a
cluster and predicted its properties. Then the crystal has been
investigated by using the Bruker FT IR spectrometer in the 400 to
4000 cm$^{-1}$ spectral range. Features observed in the spectra
unambiguously have been interpreted just as clustering of hydrogen
atoms.

\bigskip

\section{Concluding remarks}

\hspace*{\parindent}

Thus, we have uncovered that the interpretation of quantum
mechanics in the framework of the double solution theory indeed is
possible. However, the theory presented is distinguished from de
Broglie's (1987), which he actively developed seeking for the
solution of deterministic interpretation of the problem. The major
point of the given concept is an original cellular construction of
a real space, the introduction of notions of the particle, mass,
and elementary excitations of the space. The mechanics constructed
is based on the Lagrangians (16) and (23), equations of motion,
and solutions to them, (17)-(22). The Lagrangians explicitly
include elementary excitations of the space, which accompany a
moving particle and directly interact with the particle. The main
peculiarities of the mechanics called submicroscopic quantum (or
wave) mechanics are the free path lengths for the particle
$\lambda$ and its inerton cloud $\Lambda$ and, because of that,
the mechanics is similar to the kinetics theory. The particle
velocity $v_0$ is connected with $\lambda$ by relation $v_0 =
\lambda/T$ where $1/T$ is the frequency of the particle collisions
with the inerton cloud (and $1/2{\kern 0.6pt}T = \nu$ is the
frequency of the particle oscillation along its path). Since the
motion of the particle is of oscillating nature, it permits the
construction of the Hamiltonian-Jacobi equation (25) and the
obtaining the minimum increment of the particle action within the
period $\nu^{-1}$ that is identified with Planck's constant $h$.
This allows one to derive the principal quantum mechanical
relations (1) and then construct the Schr\"odinger and Dirac
formalisms.

Submicroscopic quantum mechanics has solved the spin problem reducing it to
special intrinsic pulsations of a moving particle. As a result, an
additional correction (positive or negative) is introduced to the particle's
Hamiltonian transforming it to a matrix form that in its turn has provided
the reliable background to the Dirac's linearization of the classical
relativistic Hamiltonian.

Inertons are treated as a substructure of the matter waves and yet
inertons surrounding moving particles are identified with carriers
of inert properties of the particles. The inerton concept also
determines the boundaries of employment of the wave
$\psi$-function and spinor formalisms reducing the boundaries to
the range covered by inerton cloud amplitude $\Lambda$ of the
particle studied.

At last, inertons, which widely manifest themselves in numerous
experiments, can be treated as a basis for anticipation in
physical systems because just inertons represent those inner
properties to which Dubois (2000a,b) referred constructing
anticipation as actually embedded in the systems.

Further studies need widening the scope of applying of quantum
mechanics. In particular, one could apply inertons to the problem
of quantum gravity because inertons are shown to be real carriers
of gravitational interaction. Bounias (2002) has just found other
application of inertons, namely, to biological systems: the
availability of the inerton wave function of an object allowed him
to construct the Hamiltonian of living organism considering it as
an anticipatory operator of evolution.

\bigskip

\section*{References}

\begin{enumerate}
\item% 1
Ahluwalia D. V. (1994). Quantum Measurement, Gravitation, and
Locality. \textit{Physics Letters} \textbf{B} \textbf{339},
301-303.

\item% 2
Ahluwalia D. V. (2000). Quantum Violation of the Equivalence
Principle and Gravitationally Modified de Broglie Relation,
\textit{arXiv.org e-print archive} gr-qc/0002005.

\item% 3
Amaldi U. (2000). The Importance of Particle Accelerators.
\textit{Europhysics News} \textbf{31}, 5-9.

\item% 4
Bohm D. (1952). A Suggested Interpretation of the Quantum Theory, in Terms
of "Hidden" Variables. I. \textit{Physical Review} \textbf{85}, 166-179;
(1952). A Suggested Interpretation of the Quantum Theory, in Terms of
"Hidden" Variables. II, \textit{ibid}., \textbf{85}, 180-193.

\item% 5
 Bohm D. (1996). On the Role of Hidden Variables in the Fundamental Structure
of Physics, \textit{Foundations of Physics} \textbf{26}, 719-786.

\item% 6
Bounias M. (1990). \textit{La creation de la vie: de la matiere a
l'esprit, L'esprit et la matiere}, Editions du Rocher, Jean-Paul
Bertrand Editeur.

\item% 7
Bounias M. (2000). The Theory of Something: a Theorem Supporting
the Conditions for Existence of a Physical Universe, from the
Empty Set to the Biological Self, \textit{International Journal of
Computing Anticipatory Systems} \textbf{5}, 11-24.

\item% 8
Bounias M. (2002). "The Hamiltonian of Living Organisms: an
Anticipatory Operator of Evolution." Computing Anticipatory
Systems: CASYS'01 - Fifth International Conference. International
Journal of Computing Anticipatory Systems (these volumes).

\item% 9
Bounias M., and Bonaly A. (1994). On Mathematical Links Between Physical
Existence, Observebility and Information: Towards a "Theorem of Something",
\textit{Ultra Scientist of Physical Sciences} 6, 251-259; (1996). Timeless
Space is Provided by Empty Set, \textit{ibid}. \textbf{8}, 66-71.

\item% 10
Bounias M., and Bonaly A. (1996). On Metric and Scaling: Physical
Co-ordinates in Topological Spaces, \textit{Indian Journal of
Theoretical Physics}  \textbf{44}, 303-321.

\item% 11
Bounias M., and Bonaly A. (1997). Some Theorems on the Empty Set as
Necessary and Sufficient for the Primary Topological Axioms of Physical
Existence, \textit{Physics Essays} \textbf{10}, 633-643.

\item% 12
Bounias M. and Krasnoholovets V. (2002). Scanning the Structure of
Ill-known Spaces. II. Principles of Construction of Physical
Space, to be submitted.

\item% 13
Chashihin J. (2000). To the Physical Nature of Spin. A Way to Unified Theory
of Matter, \textit{Spacetime \& Substance} \textbf{1}, no. 3, 128-131.

\item% 14
Danilov Yu. A., Rozhkov A. V., Safonov V. L. (1996). Possibility of
Geometrical Description of Quasiparticles in Solids. \textit{International
Journal of Modern Physics} \textbf{B 10}, no. 7, 777-791.

\item% 15
De Broglie L. (1960). \textit{Non-linear Wave Mechanics. A Causal
Interpretation}. Elsevier, The Netherland.

\item% 16
De Broglie L. (1986). \textit{Heisenberg's Uncertainty Relations and the
Probabilistic Interpretation of Wave Mechanics}. Mir, Moscow.

\item% 17
De Broglie L. (1987). Interpretation of Quantum Mechanics by the Double
Solution Theory. \textit{Annales de la Fondation Louis de Broglie
}\textbf{12}, no. 4, 399- 421.

\item% 18
Dubois D. M. (2000a). "Review of Incursive, Hypercursive and
Anticipatory Systems -- Foundation of Anticipation in
Electromagnetism". Computing Anticipatory Systems: CASYS'99 --
Third International Conference. Edited by Daniel M. Dubois,
Published by The American Institute of Physics, AIP Conference
Proceedings 517, pp. 3-30.

\item% 19
Dubois D. M. (2000b). "Computational Derivation of Quantum
Relativist Electromagnetic Systems with Forward-Backward
Space-Time Shifts". Computing Anticipatory Systems: CASYS'99 -
Third International Conference. Edited by Daniel M. Dubois,
Published by The American Institute of Physics, AIP Conference
Proceedings 517, pp. 417-429.

\item% 20
Dubrovin B. A., Novikov S. P., and Fomenko A. T. (1986). \textit{Modern
Geometry. Methods and Applications}, Nauka, Moscow, 1986, p. 291 (in
Russian).

\item% 21
Dunn C. (1998). \textit{The Giza Power Plant: Technologies of Ancient
Egypt}. Bear \& Company, Inc., Santa Fe.

\item% 22
Kempf A., Mangano G., and Mann R. B. (1995). Hilbert Space Representation of
the Minimal Uncertainty Relation. \textit{Physical Review} \textbf{D 52},
1108-1118.

\item% 23
Krasnoholovets V., and Ivanovsky D. (1993). Motion of a Particle and the
Vacuum. \textit{Physics Essays}, \textbf{6}, no. 4, 554-563 (also
\textit{arXiv.org e-print archive} quant-ph/9910023).

\item% 24
Krasnoholovets V. (1997). Motion of a Relativistic Particle and the Vacuum.
\textit{Physics Essays}, 10, no. 3, 407-416 (also \textit{arXiv.org e-print
archive} quant-ph/9903077).

\item% 25
Krasnoholovets V. (2000a). On the Nature of Spin, Inertia and Gravity of a
Moving Canonical Particle. \textit{Indian Journal of Theoretical Physics},
\textbf{48}, no. 2, 97-132 (also \textit{arXiv.org e-print archive}
quant-ph/0103110).

\item% 26
Krasnoholovets V. (2000b). Space Structure and Quantum Mechanics.
\textit{Spacetime \& Substance} \textbf{1}, no. 4, 172-175.

\item% 27
Krasnoholovets V. (2001a). On the Theory of the Anomalous Photoelectric
Effect Stemming from a Substructure of Matter Waves, \textit{Indian Journal
of Theoretical Physics}, \textbf{49}, no. 1, 1-32 (also \textit{arXiv.org
e-print archive} quant-ph/9906091).

\item% 28
Krasnoholovets V. (2001b). On the Way to Disclosing the Mysterious
Power of the Great Pyramid, in Research Articles on
http://gizapyramid.com.

\item% 29
Krasnoholovets V., and Byckov V. (2000). Real Inertons Against
Hypothetical Gravitons. Experimental Proof of the Existence of
Inertons. \textit{Indian Journal of Theoretical Physics},
\textbf{48}, no. 1, 1-23 (also \textit{arXiv.org e-print archive}
quant-ph/0007027).

\item% 30
Krasnoholovets V., Collective dynamics of hydrogen atoms in the
KIO$_3 \cdot$HIO$_3$ crystal dictated by a substructure of the
hydrogen atoms' matter waves, \textit{arXiv.org e-print archive}
cond-mat/0108417.

\item% 31
Leonhardt U., and Piwnicki P. (1999). Optics of Nonuniformly Moving Media.
\textit{Physical Review} \textbf{A 60}, 4301-4412.

\item% 32
Okun L. B. (1988). \textit{Physics of Elementary Particles}, Nauka, Moscow
(in Russian).

\item% 33
Plyuschay M. S. (1989). Massless Point Particle with Rigidity,
\textit{Modern Physics Letters} \textbf{A 4}, 837-847.

\item% 34
Plyuschay M. S. (1990). Relativistic Particle with Arbitrary Spin in non
Grassmannian Approach, \textit{Physics Letters} \textbf{B 248}, 299-304;
Relativistic Model of the Anyon, \textit{Physics Letters} \textbf{B 248}
107-112.

\item% 35
Plyuschay M. S. (2000). Monopole Chern-Simons Term: Charge Monopole System
as a Particle with Spin, \textit{Nuclear Physics} \textbf{B 589}, 413-439.

\item% 36
Pullin J. (1993). Knot Theory and Quantum Gravity in Loop Space: a Primer.
In \textit{Proceeding of the V Mexicon School of Particles and Fields}, ed.
J. L. Lucio, World Scientific, 1992.

\item% 37
Rangelov J. M. (2001). Physical Model of Schr\"odinger's Electron.
Heisenberg Convenient Way for Description of its Quantum
Behaviour. \textit{arXiv.org e-print archive} quant-ph/0001099;
Physical Model of Dirac Electron. Calculation of its Mass at Rest
and own Electric and Magnetic Intensities on its Momentum
Location.\textit{ arXiv.org e-print archive} quant-ph/0002019.

\item% 38
Sahni V., and Wang L. (2000). New Cosmological Model of Quintessence and
Dark Matter. \textit{Physics Review} \textbf{D 62}, 103517.

\item% 39
Schiff L. I. (1959). \textit{Quantum Mechanics}. Izdatelstvo Inostrannoi
Literatury, Moscow (Russian translation), p. 373.

\item% 40
't Hooft G. (1998). Transplanckian Particles and the Quantization of Time.
\textit{arXiv.org e-print archive} gr-qc/9805079.

\item% 41
Wallace D. (2000). The Quantization of Gravity -- an Introduction.
\textit{arXiv.org e-print archive} gr-qc/0004005.
\end{enumerate}

\end{document}